\documentclass[twocolumn,preprintnumbers,amsmath,amssymb,prb]{revtex4}

\usepackage{epsfig}
\usepackage{graphicx}
\usepackage{dcolumn}
\usepackage{hyperref}
\usepackage{bm}

\newcommand{\mean}[1]{\langle #1 \rangle}

\begin{document}

%\preprint{APS/123-QED}

\title{Statics and dynamics of a cylindrical droplet under an external body force}

\author{J. Servantie}
\author{M. M{\"u}ller}
\affiliation{Institut f{\"u}r Theoretische Physik, Friedrich-Hund-Platz 1, 37077 G{\"o}ttingen, Germany}

\date{\today}

\begin{abstract}
We study the rolling and sliding motion of droplets on a corrugated substrate
by Molecular Dynamics simulations.  Droplets are driven by an external body
force (gravity) and we investigate the velocity profile and dissipation
mechanisms in the steady state. The cylindrical geometry allows us to consider
a large range of droplet sizes. The velocity of small droplets with a large
contact angle is dominated by the friction at the substrate and the velocity of
the center of mass scales like the square root of the droplet size. For large
droplets or small contact angles, however, viscous dissipation of the flow
inside the volume of the droplet dictates the center of mass velocity that
scales linearly with the size.  We derive a simple analytical description
predicting the dependence of the center of mass velocity on droplet size and
the slip length at the substrate. In the limit of vanishing droplet velocity
we quantitatively compare our simulation results to the predictions and good
agreement  without adjustable parameters is found.
\end{abstract}

\maketitle

\section{Introduction}
Understanding and controlling the motion of droplets is crucial for many
practical applications including ink jet printing or microfluidic
devices.\cite{Quake2005} The motion of droplets can be driven in various ways:
For instance, if a temperature gradient is exerted on the fluid the droplet
will move because of the Marangoni forces caused by surface tension gradients.
\cite{Velarde1998} Alternatively, a temperature \cite{Brochard1989} or
wettability gradient \cite{Laibinis2002,Ondarcuhu1995,Thiele2004} can be
applied at the substrate.  Forces can also be exerted onto a droplet by
immersing it into a sheared, viscous fluid.\cite{Rauscher07}

Instead of transferring forces via the droplet's surface one can apply a body
force. Examples are falling drops on an inclined plane due to gravity
\cite{Pomeau1999,Pomeau2001,Thiele02} or the motion of droplets on a rotating
disk due to centrifugal forces. In these cases much attention has been devoted
to the onset of motion and the changes of the shape of the droplets and their
instabilities in response to strong body forces.\cite{Huppert82, Thiele02,
Rio05, LeGrand05, LeGrand06b} Surprisingly, the steady-state motion of droplets
under these conditions has received little attention.\cite{Kim02d} However,
this linear regime, where the droplet shape resembles its equilibrium form, is
pertinent to small droplets as their occur in nanofluidic devices. In the case
of small droplets one can ask the question by which mechanism the energy that is
imparted onto the drop via the external force is dissipated. The dominating
mechanism is expected to depend on the size of the droplet \cite{BrochardF1992} 
and it dictates the steady-state velocity of the droplet.

\begin{figure}[h!]
\centering
\includegraphics[scale=0.75]{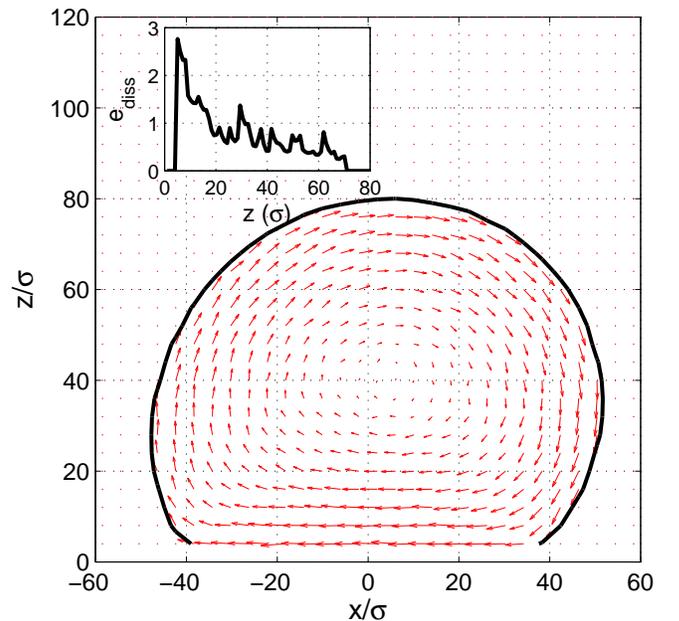}
\caption{Velocity field inside a droplet of contact angle
  $\theta_E=130^o$. The inset represents the viscous energy dissipation per
unit volume of fluid as a function of the height.}
%\caption{Snapshots of two typical droplets with respectively contact angles
%  $\theta_E=130^o$ for (a) and $\theta_E=31.5^o$ for (b).}
\label{FIGINTRO1}
\end{figure}

Similar dissipation mechanisms also act when a droplet spreads on a wettable
substrate \cite{Tanner79, Ruijter97, Ruijter99, Milchev02e, Heine03, Heine04,
Webb2005, Heine05b} but the study of the steady-state properties is
computationally much more convenient because one can average the properties
(like the velocity profiles inside the droplet shown in Fig.~\ref{FIGINTRO1}) along the
trajectory instead of averaging over many realizations of the spreading
process. 

Here we study the steady-state of cylindrical droplets under the influence of
a weak body force and determine the flow profiles inside of the droplet (see
Fig.~\ref{FIGINTRO1}) as well as the relation between the velocity of the
droplet's center of mass and its size in the linear regime.  We derive a simple
analytical approximation for the velocity inside the droplet, which is
quantitatively compared to the results of extensive Molecular Dynamics
simulations of a coarse-grain polymer model.

%----------------------------
The manuscript is organized as follows: In Sec.~\ref{sec2} we describe our
simulation model and technique.  Then, in Sec.~\ref{sec3}, we compute
properties of the bulk liquid such as surface tension and viscosity.
Secs.~\ref{sec4} and \ref{sec5} comprise the analysis of static properties,
such as contact angle and adhesion energy, and dynamics properties of droplets,
respectively. The manuscript closes with a brief discussion.

\section{The Model}
\label{sec2}

Being interested in the universal properties of liquid droplets we utilize a
coarse-grain model of a polymer liquid.\cite{kremer_grest, Bennemann99d, Muller00f, 
Muller03d} In experiments as well as simulation
studies, polymer liquids are particularly suitable because the vapor pressure
is very low and evaporation effects can be disregarded, Moreover, the chain length
allows to control the viscosity without changing the molecular
interactions. Our coarse-grained polymer model incorporates only the three
relevant interactions necessary to bring about the universal properties of
dense polymer liquids: connectivity of the segments along the extended
macromolecule, excluded volume of the segments on short distances and an
attractive interaction at intermediate distances to make the polymers condense
into a liquid. Within the framework of such a coarse-grained model, an
interaction center -- a segment -- is comprised of a small number of monomeric
repeat units.

Our model for the polymer droplets is similar to the one of Pastorino {\it et
al.} for a polymer melt interacting with a polymer brush. \cite{Pastorino2006}
The potential between neighboring monomers of a polymer is given by the Finite
Extensible Nonlinear Elastic (FENE) potential:

\begin{equation}
U_{\rm FENE}=
\left \lbrace 
\begin{array}{ccc}
-\frac{1}{2} k R_0^2 \ln \left[1-\left(\frac{r}{R_0}\right)^2\right] & {\rm for}  & r<R_0 \\
\infty & {\rm for}  & r \ge R_0
\end{array}
\right.
\label{MODELEQ1}
\end{equation}
with $R_0=1.5\ \sigma$ and $k=30\ \epsilon/\sigma^2$. The monomers additionally interact with a 6-12 Lennard-Jones potential,
\begin{equation}
U_{\rm LJ}=
\left \lbrace 
\begin{array}{ccc}
4 \epsilon \left[ \left(\frac{\sigma}{r}\right)^{12}-\left(\frac{\sigma}{r}\right)^6\right] & {\rm for} & r<r_c \\
0 & {\rm for} & r \ge r_c
\end{array}
\right.
\label{MODELEQ2}
\end{equation}
where the cutoff distance is chosen to be $r_c=2 \times 2^{1/6}\ \sigma$. The
Lennard-Jones parameters are fixed to unity, $\epsilon=1$ and $\sigma=1$. In
all the simulation the length of the polymers is fixed to $N_P=10$ monomers.

The solid, corrugated substrate is modeled by two layers of an fcc lattice of
density $\rho_s=2\ \sigma^{-3}$ interacting with the monomers via a
Lennard-Jones potential. The length scale of the potential is fixed to
$\sigma_s=0.75\ \sigma$, while the strength is varied between
$\epsilon_s=0.2-1.0\ \epsilon$.  The cutoff distance is identical to the range
of  the interactions in the fluid.  This attraction between the polymer
segments and the substrate dictates both static properties (e.g., contact
angle) and dynamic properties (e.g., slip length).

We use a dissipative particle dynamics (DPD) thermostat in our model. The
DPD thermostat has the advantage of acting locally on the system while
conserving the momentum. Hence the occurrence of non-physical 
currents in the fluid is avoided. \cite{Koelman1992,Warren1995,Kremer2003} With a DPD
thermostat two new force are added to the microcanonical equation of motion,

\begin{equation}
m \ddot{\mathbf{r}}_i=\sum_{j\ne i} \left(\mathbf{F}_{ij}+\mathbf{F}_{ij}^D+\mathbf{F}_{ij}^R\right)
\label{MODELEQ3}
\end{equation}
where $m$ denotes a segment's mass and $\ddot{\mathbf{r}}_i$ its acceleration.
$\mathbf{F}_{ij}^D$ and $\mathbf{F}_{ij}^R$ are the dissipative and fluctuating
forces acting on particle $i$ due to the interaction with particle $j$, respectively.
In order to conserve momentum these terms are applied to pairs of particles and
they take the form:

\begin{eqnarray}
\mathbf{F}_{ij}^D&=&-\gamma \omega_D(r_{ij})\left(\mathbf{1}_{ij}\cdot
  \mathbf{v}_{ij}\right)\mathbf{1}_{ij} \\
\mathbf{F}_{ij}^R&=&\zeta \omega_R(r_{ij})\theta_{ij}\mathbf{1}_{ij}
\label{MODELEQ4}
\end{eqnarray}
where  $\mathbf{1}_{ij}=\mathbf{r}_{ij}/r_{ij}$ is the unit vector pointing
from particle $j$ to particle $i$. The damping coefficient, $\gamma$, and the strength
of the noise, $\zeta$, obey the fluctuation dissipation theorem, $\zeta^2=2 k_B
T \gamma$. We fix $\gamma=0.5$ in all our simulations. The weight functions are
defined as,

\begin{equation}
\omega_R^2=\omega_D=
\left \lbrace
\begin{array}{cc}
(1-r/r_c)^2 & r<r_c \\
0           & r\ge r_c
\end{array}
\right.
\label{MODELEQ5}
\end{equation} 
and $\theta_{ij}$ in Eq.~(\ref{MODELEQ4}) is a random variable of vanishing
average and second moment: 

\begin{eqnarray}
\mean{\theta_{ij}}&=&0 \\
\mean{\theta_{ij}(t) \theta_{kl}(t')}&=&\left(\delta_{ij}
  \delta_{jl}+\delta_{il} \delta_{jk} \right) \delta(t-t')
\end{eqnarray}
We use a uniform distribution for $\theta$ because it was shown that uniform random
numbers give the same results as Gaussian random numbers. \cite{Dunweg1991} The
equations of motion are integrated with the velocity Verlet algorithm utilizing
a time step $\Delta t=0.005\ \tau$.  The temperature is set to $k_B T=1.2\
\epsilon$ in all the simulations. At this temperature the density of the fluid
is found to be $\rho_L=0.788\ \sigma^{-3}$ while the vapor density is
negligible.\cite{Muller00f, MacDowell00}  Under these conditions, the surface
tension is large enough to ensure the stability of droplets under an external
force.

\section{Bulk Properties}
\label{sec3}
In order to study the statics and dynamics of droplet one needs to evaluate
bulk properties of the fluid as a prerequisite. Quantities, such as surface
tension and shear viscosity, determine the shape of the droplet and the energy
dissipation rate, respectively. First, we compute the self-diffusion
coefficient $D$ of the polymers. The system we use in order to do that is a
cubic box of size $L=19.04\ \sigma$ containing $N=5440$ monomers.  The
diffusion coefficient of a polymer is computed from the integral of the
auto-correlation function of the velocity of its center of mass,

\begin{equation}
D=\frac{1}{3}\int_0^\infty {\rm d}t \; \mean{\mathbf{v}_{\rm
    CM}(t)\cdot\mathbf{v}_{\rm CM}(0)}
\label{BULKEQ1}
\end{equation}
We find a self-diffusion coefficient $D=0.0157 \pm 0.0030 \ \sigma^2/\tau$.
Moreover, computing the end-to-end distance of a polymer we obtain
$R_{ee}=\sqrt{\mean{R_{ee}^2}}=3.447\ \sigma$. The Rouse relaxation time of the
polymer is $\tau_P=R_{ee}^2/(3\pi^2 D)=25.6 \pm 5\ \tau$. \cite{DoiEdwards}
This relaxation time permits to estimate an upper limit of the shear rate that one
can apply to the fluid before the polymers starts to be significantly deformed:
The Weissenberg number ${\rm We}=\dot{\gamma}\tau_P$ should be of the order of
unity or less, hence $\dot{\gamma}_{\rm max}\approx 0.05\ /\tau$. 

The shear viscosity, $\eta$, is obtained from the auto-correlation function of
the off-diagonal elements of the microscopic flux associated to viscosity,
\cite{McQuarrie} 

\begin{equation}
\eta=\frac{1}{k_B T V} \int_0^\infty dt \mean{J_{\alpha \beta}(t)  J_{\alpha \beta}(0)}
\label{BULKEQ2}
\end{equation}
where $V$ is the volume of the fluid and the microscopic flux is,
\begin{equation}
J_{\alpha \beta}=\left(\sum_i m v_i^\alpha
  v_i^\beta+\sum_{i<j}F_{ij}^\alpha r_{ij}^\beta \right)=V P_{\alpha \beta}
\label{BULKEQ3}
\end{equation}
where $P_{\alpha \beta}$ denotes the pressure tensor. We use the same system
parameters as for the calculation of the diffusion coefficient to compute the
shear viscosity. Averaging over $2\cdot 10^5$ time units $\tau$ yields the
value $\eta=5.3 \pm 0.1\ \sigma^2/\sqrt{m \epsilon}$.

Finally, we compute the surface tension, $\gamma$, using a slab geometry.
\cite{Sikkenk1988,Alejandre2002} In this case the surface tension is found from
the anisotropy of the pressure tensor according to:

\begin{equation}
\gamma=\frac{1}{2 L_y L_z} \left[\mean{J_{xx}}-\frac{1}{2}\left(\mean{J_{yy}}+\mean{J_{zz}}\right)\right]
\label{BULKEQ4}
\end{equation}
where $L_y=L_z=30\ \sigma$ are the lengths of the simulation box in the $y$ and
$z$ directions, respectively. Thus, $2 L_y L_z$ is the surface area of the
fluid. The box size is $L_y=L_z=30\ \sigma$ and $L_x=100\ \sigma$.  In order to
compute $\gamma$ we use a larger system than previously, namely $N=20\,000$
monomers. The value of the surface tension we obtain is $\gamma=0.5154\pm0.0023
\ \epsilon/\sigma^2$.

\section{Statics}
\label{sec4}

\begin{figure}[h!]
\centering
\includegraphics[scale=0.6]{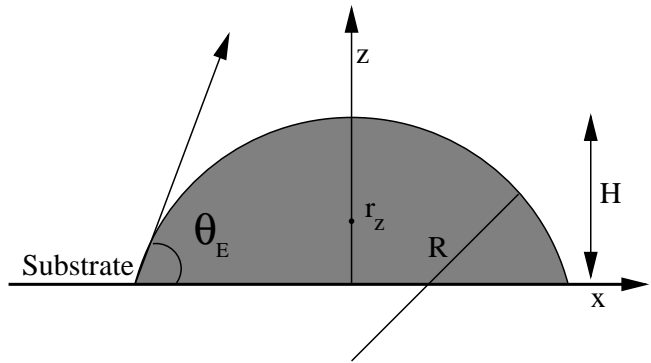}
\caption{Sketch of a droplet defining the notations in the text.}
\label{STATICSFIG1}
\end{figure}

In this section, we compute the contact angle and adhesion energy of droplets
for various strengths of the polymer-substrate interaction, $\epsilon_s$, and
system sizes, $N$. We use periodic boundaries along the $y$ axis and the droplet
spans the simulation cell in this direction. As a consequence, the droplets are
cylindrical. By virtue of the cylindrical shape, the radius of the droplet
increases as $\sim \sqrt{N}$ (instead of $N^{1/3}$ for a cap-shaped drop)
and, thus, larger droplets can be studied.  The size of the simulation cell
along the $y$ axis is fixed to $L_y=19.93 \ \sigma$ in all the following
simulations. One can characterize the geometry of a cylindrical droplet by
two parameters -- the equilibrium contact angle, $\theta_E$, and the radius,
$R$.  We have schematically depicted a droplet in Fig.~\ref{STATICSFIG1}. 

\begin{figure}[h!]
\centering
\includegraphics[scale=0.75]{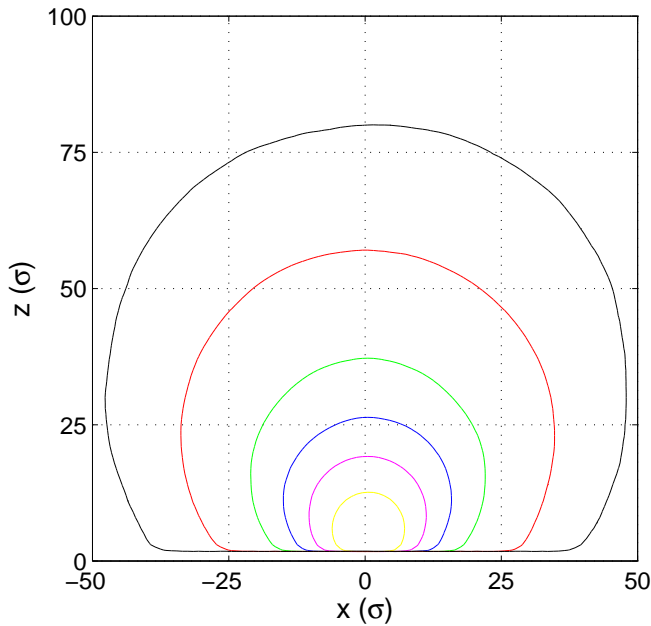}
\caption{Density contours for increasing system sizes varying from 
$N=2000$ to $N=100000$  monomers for $\epsilon_s=0.4\ \epsilon$. 
The equilibrium contact angles is $\theta_E=130^o$.}
\label{STATICSFIG2a}
\end{figure}

\begin{figure*}
\centering
\includegraphics[width=18cm]{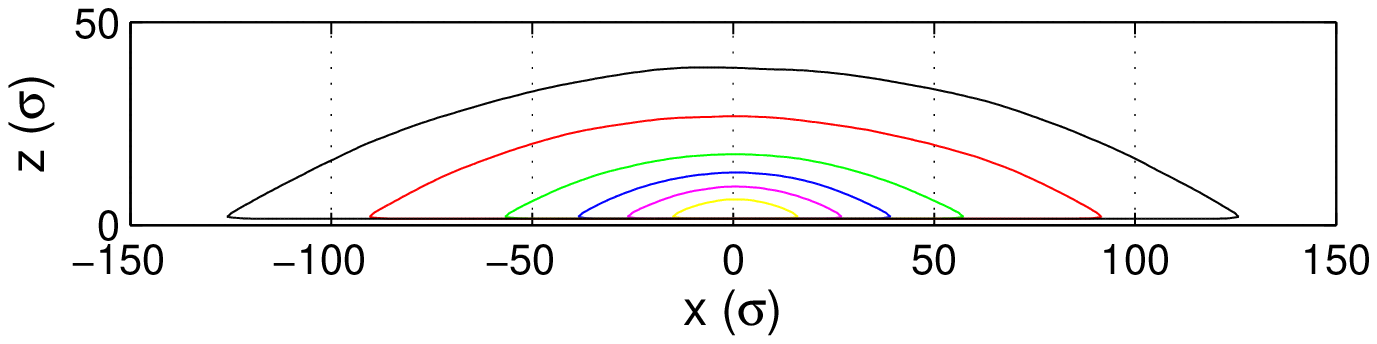}
\caption{Density contours for increasing system sizes varying from 
$N=2000$ to $N=100000$  monomers for $\epsilon_s=0.8\ \epsilon$. 
The equilibrium contact angles is $\theta_E=29.4^o$.}
\label{STATICSFIG2b}
\end{figure*}

Other quantities such as volume $V$, area $A$ of contact with the
substrate, and height of the center of mass $r_z$ can be expressed as a
function of $R$ and $\theta_E$. For a cylindrical droplet one obtains the
following relations:

\begin{eqnarray}
V&=&\frac{R^2}{2}\left(2\theta_E-\sin 2 \theta_E \right) L_y \label{STATVOL}\\
A&=&2 R \sin \theta_E L_y \label{STATAREA}\\
r_z&=&R \left( \frac{4}{3} \frac{\sin^3 \theta_E}{2\theta_E-\sin 2\theta_E}-\cos\theta_E\right) \label{STATRZ}
\end{eqnarray}
We have computed the equilibrium density profiles for system sizes ranging from
$N=2\,000$ to $N=100\,000$ monomers and substrate strengths varying between
$\epsilon_s=0.2\ \epsilon$ and $\epsilon_s=0.8\ \epsilon$. The statistical
average of the density is typically extended over $20\,000$ time units $\tau$
and the spatial binning size for the density profiles is chosen as $1\ \sigma$
in both directions.  The results are depicted in Figs.~\ref{STATICSFIG2a} and
\ref{STATICSFIG2b}.


\begin{figure}[h!]
\centering
\includegraphics[scale=0.6]{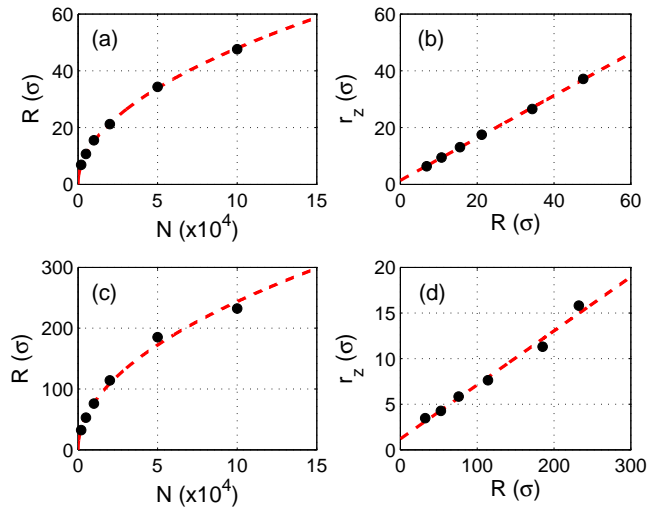}
\caption{Radius as a function of the number of monomers and the center 
mass height as a function of the radius for $\epsilon_s=0.4\ \epsilon$ in 
respectively (a) and (b), and for $\epsilon_s=0.8\ \epsilon$ in (c) and
(d). The circles are the results of the simulations and the dashed line the fits.}
\label{STATICSFIG3}
\end{figure}

The contour line represents the average density, $(\rho_L+\rho_V)/2$, between
the density of the dense polymer liquid, $\phi_L$ and the vapor, $\rho_V$ which
coexists with it.  We note that the vapor density is very small ($\sim 10^{-6}\
\sigma^{-3}$). Utilizing the relation between center of mass height and contact
angle (\ref{STATRZ}), one can compute the contact angle in a systematic way
without referring to the detailed shape of the liquid-vapor interface in the
vicinity of the three-phase contact line.  Indeed, one can easily fit a circle
to the contours of Figs.~\ref{STATICSFIG2a} and \ref{STATICSFIG2b} and obtain
the curvature, $R$, of the droplet. On the other hand, the height, $r_z$, of
the center of mass is straightforwardly monitored in the course of the
simulation.  We depict in Fig.~\ref{STATICSFIG3} the dependence of the radius
on the number of monomers, $N$, and the height of the center of mass, $r_z$, as a function of
the radius for both systems. As we see, $r_z$ linearly increases with $R$
showing that there is no significant dependence of the contact angle on size.
First, the length of the three-phase contact line is independent of the droplet
size for a cylindrical droplet. \cite{Kulmala2007,Neumanna1998} Second, the
interface potential, which describes the interaction between the liquid-vapor
interface and the substrate, is rather short-ranged. Third, by virtue of the
low temperature and the concomitantly low vapor density and low liquid
compressibility, the shift of the pressure away from the bulk coexistence value
due to the Laplace pressure of the curved liquid-vapor interface of the droplet
does not give rise to a noticeable change of the liquid or vapor properties.

The slopes of Figs.~\ref{STATICSFIG3}(b) and \ref{STATICSFIG3}(d) yield the
equilibrium contact angles $\theta_E=130^o$ and $\theta_E=29.4^o$ for
$\epsilon_s=0.8 \epsilon$ and $\epsilon_s=0.4 \epsilon$, respectively. 

We observe that $r_z$ does not vanish for zero radius for two reasons: First,
the second layer of the substrate is approximately located at a height of $0.7\
\sigma$. Second, there is a depletion layer of the polymer liquid near the
substrate that shifts the height of the center of mass farther upwards.  We
applied this systematic method to other substrate strengths in order to
evaluate how the contact angle and the adhesion free energy depends on
$\epsilon_s$. In equilibrium the macroscopic contact angle of a droplet point
is determined by the Young equation, \cite{Young,deGennes1985}

\begin{equation}
\gamma_{SV}-\gamma_{SL}-\gamma \cos \theta_E=0
\label{STATICSEQ1}
\end{equation}
where $\gamma_{SV}$, $\gamma_{SL}$ denote the surface energies between the
substrate and the vapor phase, and the substrate and the liquid phase,
respectively. One can rewrite Eq.~(\ref{STATICSEQ1}) in terms of the adhesion
free energy per area unit $W$ as,

\begin{equation}
W=\gamma \left(1+\cos \theta_E \right).
\label{STATICSEQ2}
\end{equation}

As a consequence, the fluid will completely wet the surface if $W\ge 2\gamma$.
When $W< 2\gamma$, however, there will be only partial wetting, i.e., droplets
with a finite contact angle will form. Hence, by representing $W$ as a function
of the strength of the substrate, $\epsilon_s$, one can extrapolate towards the
value where the wetting transition occurs. 

\begin{figure}[h!]
\centering
\includegraphics[scale=0.6]{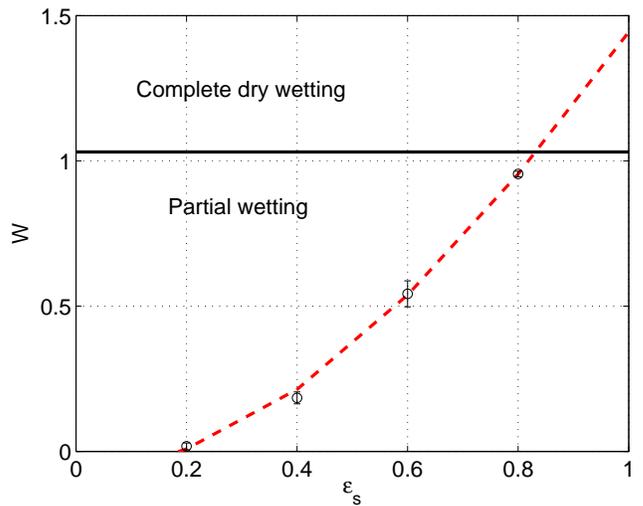}
\caption{Adhesion free energy $W$ per area units versus the strength of the substrate
  $\epsilon_s$. The circles are the data obtained from contact angle
  computations and with Eq. (\ref{STATICSEQ2}) while the dashed line is from the average Lennard-Jones
  potential energy according to Eq. (\ref{STATICSEQ5}).}
\label{STATICSFIG4}
\end{figure}

Alternatively, one can compute the adhesion energy from the Molecular Dynamics
simulations by computing the average energy of the polymer-substrate
interactions, $\mean{U_{LJ}^s}$.  Utilizing the statistical mechanics
definition of the free energy

\begin{equation}
F=-k_BT\ \ln Z
\qquad \mbox{where} \;
Z=\int d\Gamma e^{-\beta \epsilon_s V_s},
\label{STATICSEQ4}
\end{equation}
denotes the partition function and $U_{LJ}^s=\epsilon_s V_s$ the Lennard-Jones
interaction between polymer segments and substrate. We differentiate
Eq.~(\ref{STATICSEQ4}) with respect to $\epsilon_s$. This procedure  yields
$\partial F/\partial \epsilon_s=\mean{V_0}=\mean{U_{LJ}^s}/\epsilon_s$.
Integration gives us the free energy as a function of the strength of the
polymer-substrate interaction:

\begin{equation}
F=\int_0^\epsilon d\epsilon' \frac{\mean{U_{LJ}^s}}{\epsilon'}+C.
\label{STATICSEQ5}
\end{equation}
where $C$ is an unknown constant of integration. The adhesion energy $W$ is
then simply, $W=\frac{F}{A}$, where $A$ is the area of the interface between
the fluid and the substrate.
We
have computed this adhesion energy for values of the strength between
$\epsilon_s=0.2-1\ \epsilon$. The fluid is confined between two substrates
perpendicular to the $z$ direction while periodic boundary conditions are
applied along the $x$ and $y$ directions. The system dimensions are $L_x=19.84\
\sigma$ and $L_y=19.93\ \sigma$ while the distance between the substrates
varies between $L_z=19.74-20.96\ \sigma$ in order to maintain the coexistence
density in the fluid sufficiently far away from the substrate (bulk). The
system is comprised of N=5\,440 monomers. In Fig.~\ref{STATICSFIG4} we present
the results obtained by both methods.

We see that both methods to compute the adhesion free energy are in rather good
agreement. The constant of integration is found to be $C\approx -0.05$. This
suggest the presence of a weak first order or possibly second order drying
transition for small values of $\epsilon_s$ in accord with previous studies of
Lennard-Jones monomer fluids \cite{Adams91, Vanswol91, Bruin95} and polymeric
fluids at a higher temperature.\cite{Muller00i,Muller03d} The small value of
the adhesion energy for small values of $\epsilon_s$ allows the droplet to
easily detach from the substrate due to thermal fluctuations. If 
an external body force is applied, driven droplets will be unstable under these
conditions.  On the other hand, the extrapolation of the data for large values
of $\epsilon_s$ shows that a wetting transition of first order occurs at a
substrate strength of approximately $\epsilon_s^{\rm wet}=0.82\ \epsilon$. 

The fact that $W$ is a non-linear function of the substrate strength,
$\epsilon_s$, can be rationalized by writing the total substrate-fluid
potential energy in the form

\begin{equation}
U_{\rm TOT}=\int_0^{r_c} dz \rho(z) U_{\rm wall}(z).
\label{STATICSEQ7}
\end{equation}
While the substrate potential scales like $U_{\rm wall} \sim \epsilon_s$, the
density also increases with $\epsilon_s$ because the wall becomes increasingly
attractive giving rise to a strong increase of the adhesion energy and the
layering at the substrate.

\section{Dynamics}
\label{sec5}

\subsection{Boundary condition}

When a fluid confined between two walls is sheared with a shear rate
$\dot{\gamma}$ a no-slip boundary condition is often observed.\cite{Bocquet07c,Lauga07}
However, a finite slippage velocity can occur at non-wetting substrates.
\cite{Troian1997,Bocquet1999} The simplest boundary condition that can describe the
fluid velocity at the substrate is the Navier slip condition\cite{Navier1823}

\begin{equation}
\left. \eta \frac{\partial v_x}{\partial z} \right|_{z=z_h}= \lambda {v_s}
\label{DYNAEQ1}
\end{equation}
that quantifies the balance between the viscous stress due to the shear in the
fluid and the friction stress at the boundary. $\eta$ denotes the viscosity of
the liquid, and $\lambda$ is the friction coefficient.  $v_s=v_x(z_h)$ denotes
the slip velocity at the hydrodynamic position, $z_h$ of the substrate. 
The quantity $\delta$ 

\begin{equation}
\delta=\frac{\eta}{\lambda}
\label{DYNAEQ5}
\end{equation}

characterizes the slip length.  If the shear rate $\dot{\gamma}$ is not too
large then the slip length, $\delta$, does not depend on $\dot{\gamma}$  and it
can be computed by non-equilibrium molecular dynamics simulation. To this end,
one regards the velocity profiles obtained from Couette flow, (i.e.,
boundary-driven planar shear flow) and Poiseuille flow (i.e., a bulk force is
applied to the fluid).

\begin{figure}[h!]
\centering
\includegraphics[scale=0.6]{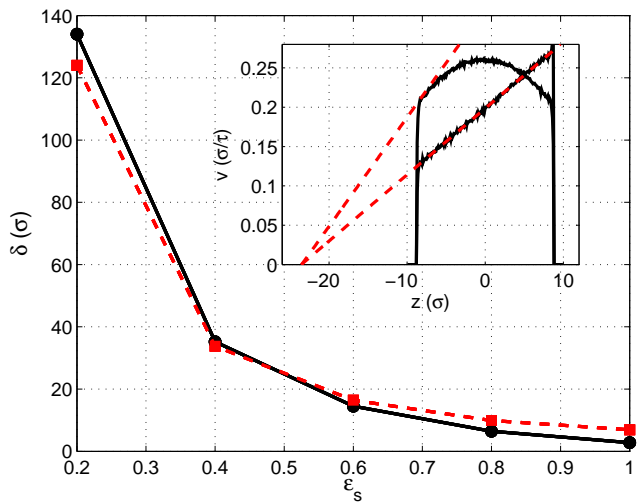}
\caption{Slip length $\delta$ as a function of the substrate strength 
$\epsilon_s$. The plain line is obtained from the Couette and Poiseuille 
profiles while the dashed line from the Green-Kubo relation (\ref{DYNAEQ4}).
The inset presents the velocity profiles of the Couette and Poiseuille flow
from which the slip length is estimated for $\epsilon_s= 0.6\ \epsilon$.
}
\label{FIGBOUND1}
\end{figure}

The linear velocity profile of Couette flow,

\begin{equation}
v_x(z)=\frac{v_s}{\delta}\left[z-z_h-\delta\right]
\label{DYNAEQ2}
\end{equation}
yields the parameter combination $z_h+\delta$. Additionally, using the velocity
profile of Poiseuille flow 

\begin{equation}
v_x(z)=\frac{\rho_l a}{\eta} \left[z_h
  \delta+\frac{1}{2}z_h^2-\frac{1}{2}z^2\right]
\label{DYNAEQ3}
\end{equation}
one can compute $z_h$ and $\delta$, independently. Alternatively, one can also
evaluate $\delta$ from equilibrium molecular dynamics by integrating the
auto-correlation function of the tangential force exerted by the wall on the
fluid, \cite{Bocquet1993,Bocquet1994}

\begin{equation}
\lambda=\frac{1}{k_B T A} \int_0^{\infty} dt \mean{F_s(t) F_s(0)}
\label{DYNAEQ4}
\end{equation}
and then the slip length is simply given by Eq.~(\ref{DYNAEQ5}) where $\eta$ is
the shear viscosity of the fluid in the vicinity of the substrate.  The position
$z_h$ of the interface is about $\sigma$, which agrees with the shift due to the
presence of the substrate. In Fig.~\ref{FIGBOUND1} we show in
the slip length computed by both the non-equilibrium
method and the equilibrium one.

For strong attraction, $\epsilon_s$, between polymer and substrate the
Green-Kubo method yields slightly larger values of the slip length than the
non-equilibrium method. This is partially due to the fact that we employ the
value of the shear viscosity corresponding to the (bulk) coexistence density,
$\rho_L$, of the liquid, while there are important layering effects present at
the substrate for large $\epsilon_s$,  enhancing the effective viscosity at the
substrate.  The largest slip length we observe is smaller than $200\ \sigma$
compatible with a recent experiment suggesting that slip lengths are of the
order of nanometers. \cite{Bocquet2006}

\subsection{Steady state velocity}

\begin{figure}[h!]
\centering
\includegraphics[scale=0.5]{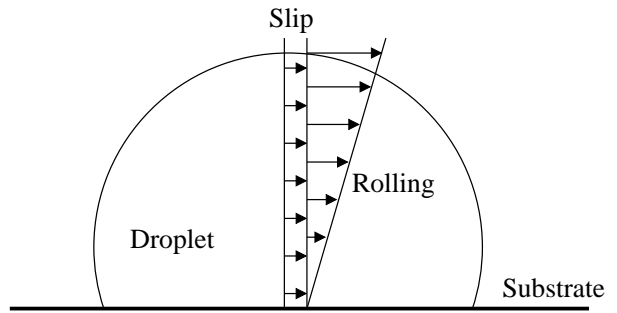}
\caption{Sketch of the velocity field inside the droplet.}
\label{DISSFIG1}
\end{figure}

In order to study the dynamics of the droplet, we apply a constant acceleration
$a$ to all particles in the system. After a relaxation time, the droplet
attains a steady state where the energy imparted onto the system by the body
force is dissipated by the viscous liquid. According to de Gennes
\cite{deGennes1985} the total energy dissipation rate in the droplet can be
decomposed in the following way:

\begin{equation}
T\sum_{\rm TOT}=T\left(\sum_w+\sum_l+\sum_f\right)
\label{DISSEQ1}
\end{equation}
where $\sum_w$ is the viscous dissipation due to the rolling motion,
$\sum_{\ell}$ denotes the dissipation at the three-phase contact line, and
$\sum_f$ denotes the dissipation in the precursor film. In our system, the
third term can be discarded because there is no precursor film. The second
term is due to a possible pinning of the particles of the fluid to the
substrate enhancing dissipation at the contact line largely. However, since we
use a regularly and finely corrugated substrate in our study one cam assume that such
processes do not play an important role. Thus, the total dissipation should
simply be:

\begin{equation}
T\sum_{\rm TOT}=T\sum_w
\label{DISSEQ2}
\end{equation}
However, an extra term should be added to Eq.~(\ref{DISSEQ2}) in case of a finite
slip, $\delta>0$, at the substrate. As sketched in Fig.~\ref{DISSFIG1}, the velocity field inside the
droplet can be decomposed into a rolling motion and the uniform translation with
the slippage velocity. This slippage at the substrate results in a second
source of dissipation. Due to the corrugation of the substrate, the fluid
in its vicinity undergoes important velocity gradients enhancing the dissipation largely. 
This description is the microscopic view of the friction between liquid and substrate.
Thus, the total dissipation rate can be written as

\begin{equation}
T\sum_{\rm TOT}=T\left(\sum_w+\sum_A\right),
\label{DISSEQ3}
\end{equation}
where the term $\sum_A$ is the friction dissipation and it is proportional to
the area of contact $A$ between the droplet and the substrate.  The second term
on the right hand side is exactly known: The power dissipated by the substrate
is 
\begin{equation}
T\sum_A=F_k v_s
\label{DISSEQ4}
\end{equation}
where the dynamic friction force $F_k$ is, \cite{Bocquet1994}

\begin{equation}
F_k=\eta \frac{v_s}{\delta} A
\label{DISSEQ5}
\end{equation}

\begin{figure}[h!]
\centering
\includegraphics[scale=0.6]{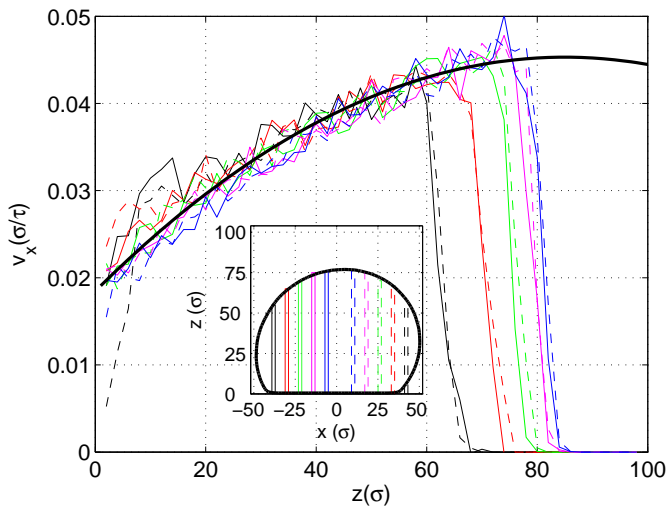}
\caption{Velocity profile as a function of the height $z$ for different lateral
positions, $x$, in the droplet. The inset shows the sections which have been
used to calculate the velocity profile $v_x(x,z)$.}
\label{DISSFIGNEW}
\end{figure}

If the droplet is sufficiently flat, the dissipation due to the rolling motion
can be evaluated in the framework of the lubrication approximation, 

\begin{equation}
T\sum_w=\eta \int_V dV \left(\frac{\partial v_x}{\partial z}\right)^2
\label{DISSEQ6}
\end{equation}
where $V$ denotes the volume of the droplet. The viscous dissipation scales like
the volume $V$ times the square of the shear rate. Thus, for flat droplets, one
obtains

\begin{equation}
T\sum_w=\xi \eta \left(\frac{v_s}{\delta}\right)^2 V
\end{equation}
where $\xi$ is a numerical factor that describes the flow profile inside the
droplet and that only depends on the contact angle. Here, we assume that the
shear at the substrate $\frac{\partial v_x}{\partial z}|_{z=z_h}=v_s/\delta$ sets the
scale of the shear inside the entire droplet. This assumption is obviously
fulfilled for droplets with a small contact angle and a large slip length.  We
will explain below that this viscous volume dissipation is most important for
flat drops. Moreover, we shall verify below that this crude assumption remains
a reasonably good description even for droplets with a larger contact angle
(cf.~Fig.~\ref{DISSFIGNEW}). 

Now, using the fact that the total energy dissipation rate of the system is simply
the velocity of the center of mass times the bulk force, one arrives at

\begin{eqnarray}
\rho_L V a U &=&\eta \frac{v_s^2}{\delta} A+\xi \eta
\left(\frac{v_s}{\delta}\right)^2 V 
\label{DISSEQ12}\\
&\approx &\eta \left(\frac{U}{\delta+r_z}\right)^2 (\delta A+\xi V)
\end{eqnarray}
where, in the last step, we have approximated the scale of the typical shear inside the droplet 
by $U/(\delta+r_z)$.

Eq.~(\ref{DISSEQ12}) describes the cross-over between two regimes: For small
drops, $\delta A \gg V$ or $\delta \gg r_z$, the friction at the substrate is
the dominant dissipation mechanism. In this limit, we obtain $\eta U/\rho_La
\sim V \delta^2/\delta A \sim R \sim \sqrt{N}$.  In the opposite limit,
the viscous dissipation inside the droplet dominates and $\eta U/\rho_La \sim
R^2 \sim N$. This phenomenological argument can be generalized to
cap-shaped droplets in three dimensions. Then, the rational yields $U/\rho_La
\sim R \sim N^{1/3}$ and $U/\rho_La \sim R^2 \sim N^{2/3}$ for small and large droplets,
respectively.

To turn this phenomenological consideration into a more quantitative description
some information about the velocity field inside the droplet is required. The
flow inside the droplet is governed by the steady-state Navier-Stokes equation
which in the limit of small Reynolds number takes the particularly simple form:
\begin{equation}
\frac{\partial^2 v_x}{\partial z^2}=-\frac{\rho_L a}{\eta}
\label{DISSNS}
\end{equation}
This assumption is justified if the velocity $U$ of the droplet is sufficiently small.
Thus, the profile of the velocity $v_x$ parallel to the substrate is a parabolic
function of the coordinate, $z$, perpendicular to the substrate\cite{deGennes1985}
\begin{equation}
v_x=\alpha z^2+\beta z +\gamma
\label{DISSEQ7}
\end{equation}
We additionally assume that the coefficients, $\alpha$, $\beta$, and $\gamma$ do 
not depend on the lateral position, $x$, inside the drop. From Eq.~\ref{DISSNS} we
read off the first coefficient, $\alpha = - \frac{\rho_L a}{2\eta}$. 

Second, we require the tangential stress to vanish at the top of the droplet,
i.e., at $z=H$, because the vapor has vanishingly small density and viscosity.
By virtue of the symmetry of the (static) droplet profile, the tangential
velocity coincides with $v_x$ at the top.

\begin{equation}
\left. \frac{\partial v_x}{\partial z}\right|_H=0
\end{equation}
This condition determines the coefficient $\beta$ and one obtains
\begin{equation}
v_x=\frac{\rho_L a}{\eta}\left(H-\frac{z}{2}\right)z+\gamma
\end{equation} 

Finally, we utilize the Navier slip condition (cf.~Eq.~(\ref{DYNAEQ1})) at the
substrate. In the following we set the origin of the $z$ axis such that $z_h=0$.

\begin{equation}
\left. v_x(z)\right|_{z_h=0}=v_s=\gamma = \delta \left. \frac{\partial v_x}{\partial z} \right|_{z_h=0} = \frac{\rho_La}{\eta}H
\end{equation}

This equation determines the slip velocity at the substrate
\begin{equation}
v_s=\frac{\rho_L a}{\eta}H \delta
\label{DISSEQ10}
\end{equation}
and the velocity profile takes the final form
\begin{equation}
v_x=\frac{\rho_L a}{\eta}\left[\left(H-\frac{z}{2}\right)z+\delta H\right]
\label{DISSEQ9}
\end{equation} 

Eq.~(\ref{DISSEQ9}) provides an approximation of the velocity profile inside
the droplet as a function of the droplet shape, the slip length, $\delta$, and
the force density $\rho_La$. 

We emphasize that the height appearing in Eq.~(\ref{DISSEQ9}) $H=R(1-\cos
\theta_E)$ does not depend on the lateral position, $x$.  This is in marked
contrast to previous work,\cite{deGennes1985} where $H$ was chosen as the local
height $h(x)$ of the droplet, which characterizes its shape. This
identification, however, would result in a spurious singularity of the viscous
dissipation at the three-phase contact line.\cite{Huh71,Kim02b} Note that the maximal velocity in
$x$ direction, $v_x^{\rm max}=v_x|_{z=h(x)}$, decreases with the lateral
distance from the center of the drop. This behavior implicitly captures some
aspects of the rolling motion (cf.~Fig.~\ref{FIGINTRO1}).

All parameters that dictate the velocity profile in Eq.~(\ref{DISSEQ9}) have
been independently determined: The viscosity $\eta$ and coexistence density
$\rho_L$ are known from the bulk properties of the fluid, the contact angle
$\theta_E$ from static considerations of droplets and the slip length $\delta$
from the Poiseuille and Couette profiles. One can thus predict the velocity
profile and quantitatively compare them to the results of the molecular
dynamics without any adjustable parameter. This comparison is presented in
Fig.~\ref{DISSFIGNEW}.

The comparison validates the basic assumption that the lateral velocity $v_x$
is independent of the lateral position, $x$. Indeed, the simple analytical
expression (\ref{DISSEQ9}) provides an excellent description of the simulation 
data using the independently determined parameters.

Considering the velocity at the top of the droplet $v_x=v_x(H)$ one
can evaluate the crossover radius $R_c$ between surface friction and viscous
volume dissipation regimes as

\begin{equation}
R_c=\frac{2\delta}{1-\cos\theta_E}
\end{equation}
Thus in order to have a large crossover radius, $R_c$, one should have
a large slip length $\delta$ and/or a small contact angle, $\theta_E$.

The detailed velocity profile inside the droplet may be difficult to measure
experimentally but the cross-over between the two regimes can be observed by
the steady-state velocity of the center of mass. Approximating the center of
mass velocity simply by the velocity $v_x(r_z)$ at the height of the center of
mass $r_z$, we obtain:

\begin{equation}
U=\frac{\rho_L a}{\eta}\left[\left(H-\frac{r_z}{2}\right)r_z+\delta H\right]
\label{DISSEQ14}
\end{equation}

In order to test our model we computed the dependence of the steady-state
velocity on the droplet size for two substrate strengths, $\epsilon_s=0.4\
\epsilon$ and $\epsilon_s=0.8\ \epsilon$. We first note that our model is valid
only for small Reynolds numbers such that no inertial effects are present and the
droplet shape resembles its equilibrium shape. Moreover, in this linear regime
the center of mass velocity is proportional to the acceleration and, thus, we
plot the ratio $U/a$ instead of $U$. 

Care must be taken in choosing the acceleration, $a$. A too small value of
$a$ gives rise to impractically long computation times and large statistical
errors in the velocity profiles and the average center of mass velocity. If one
chooses $a$ too large, however, non-linear effects become important and one
tends to underestimate the ratio $U/a$.  Since at a fixed acceleration, $a$,
the velocity of center of mass increases with size, one reaches the non-linear
regime when the droplet size grows.  We note that at very high velocities two
different behavior can be observed: For small contact angles, the droplets
break up into smaller pieces, while for large contact angles they detach from
the substrate. 

\begin{figure}[h!]
\centering
\includegraphics[scale=0.6]{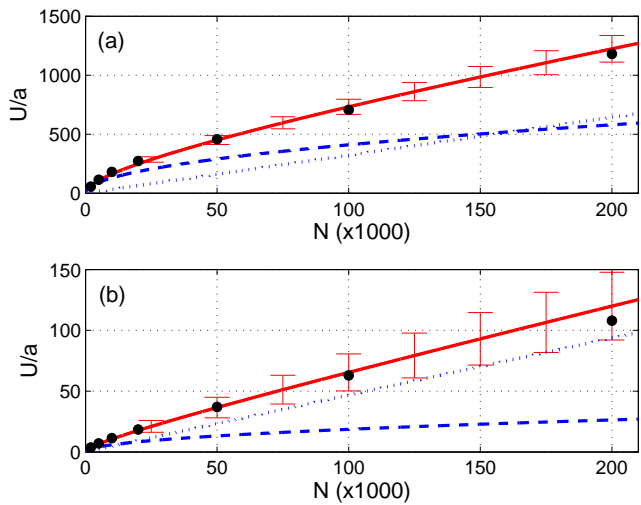}
\caption{Velocity of the center of mass per unit acceleration as a 
function of the number of monomers for (a) $\epsilon_s=0.4\ \epsilon$ 
and (b) $\epsilon_s=0.8\ \epsilon$. The circles are the results of the
molecular dynamics simulations and the dashed lines the analytical results.
The error bars mark the uncertainties associated with the independently
determined input parameters. The asymptotic limits for small and large
droplets are indicated by the solid lines proportional to $\sqrt{N}$ 
and $N$, respectively.}
\label{FIGDYNA1}
\end{figure}

\begin{table}
\begin{tabular}{lccr}
\hline
\hline
$\epsilon_s/\epsilon$ & $N$ & $U \tau/\sigma$ & $a \tau^2/\sigma$ \\
\hline
0.4 & 2000    & 0.0056    &  0.0001 \\ 
0.4 & 5000    & 0.0114    &  0.0001 \\
0.4 & 10000   & 0.0180    &  0.0001 \\
0.4 & 20000   & 0.0272    &  0.0001 \\
0.4 & 50000   & 0.0230    &  0.00005 \\
0.4 & 100000   & 0.0364    &  0.00005 \\
0.4 & 200000   & 0.0295    &  0.000025 \\
0.8 & 2000    & 0.000367    & 0.0001  \\ 
0.8 & 5000    & 0.000699    & 0.0001  \\
0.8 & 10000   & 0.00114    &  0.0001 \\
0.8 & 20000   & 0.00185    &  0.0001 \\
0.8 & 50000   & 0.00185    &  0.00005 \\
0.8 & 100000   & 0.00162    &  0.000025 \\
0.8 & 200000   & 0.0027    &  0.000025 \\
\hline
\hline
\end{tabular}
\caption{Parameters and results of the Molecular Dynamics simulations of 
droplets driven by an external body force.}
\label{tab1}
\end{table}

In Fig.~\ref{FIGDYNA1} and Tab.~\ref{tab1} we present the results of the
simulations and compared them to the predicted behavior (\ref{DISSEQ14}).  We
observe that Eq.~(\ref{DISSEQ14}) is in excellent agreement with the results of
the Molecular Dynamic simulations. The error bar on the prediction characterize
the uncertainty associated with the error of the input parameters. The figure
also depicts the asymptotic limits for small and large droplets. Our simulation
data are clearly in the cross-over region. The data for the droplet with the
larger contact angle (cf.~panel a) are mainly in the surface friction
dominated regime, while the flatter droplets (shown in panel b) already
reach the regime of volume dissipation. The
crossover between surface dominated dissipation and volume dominated
dissipation occurs at $N_c \approx 160\,000$ and $N_c \approx 33\,000$ monomers
for $\epsilon_s=0.4\ \epsilon$ and $\epsilon_s=0.8\ \epsilon$, respectively.
Consequently, surface effects are more important than volume dissipation in the
droplet with the large contact angle. This can be rationalized as follows: In
droplets with large contact angles the velocity profile in the upper part of
the droplets is close to a purely rotational motion which is characterized by
very low dissipation.  \cite{Pomeau1999} This is clearly seen in
Fig.~\ref{FIGINTRO1} which depicts the velocity profile in the comoving frame
for a droplet with $N=100\,000$ monomers and a bulk acceleration $a=0.00005\
\sigma/\tau^2$. Thus, the viscous dissipation in the inside of the droplet is
small although the volume is large.

This explanation is corroborated by the profile of the viscous energy
dissipated per unit volume of the fluid. In the inset of Fig.~\ref{FIGINTRO1}
we show this quantity as a function of the height, $z$. 
Unfortunately, we are unable to calculate the analog quantity for the droplet
with the smaller contact angle with our computational resources because for
flat droplets the linear regime is limited to very small accelerations and
the velocity of the center of mass is orders of magnitudes smaller than
the thermal velocity, $v_T = \sqrt{k_BT/m}= 1.095 \, \sigma/\tau$.

As we see in the Fig.~\ref{FIGINTRO1} the energy dissipation rate in a droplet
with a large contact angle is twice as large in the lower region than it is in
the upper one where the stream lines resemble those of a purely rotational
flow. On the other hand, if we decrease the contact angle, the flow inside the
droplet will increasingly deviate from this Huygens motion and, consequently
the region where important viscous dissipation occurs will expand.  As a
result, for a small droplet of fixed volume, if the area of contact with the
substrate is large ($\theta_E$ small) volume dissipation dominates, while if
the area of contact is small ($\theta_E$ large) surface dissipation dominates.

\section{Conclusion}

In this paper we studied equilibrium and non-equilibrium properties of polymer
droplets in contact with a corrugated substrate by Molecular Dynamics
simulations and quantitatively compared the results to a simple analytical
model. First, we measured the contact angles for different substrate strength
via a geometrical method that does not rely on the detailed shape of the
liquid-vapor profile at the three-phase contact line. Additionally, we
evaluated the adhesion free energy of the droplet and found good agreement
between the shape of the cylindrical droplets and the macroscopic contact angle
predicted by the Young equation for our coarse-grain polymer model.  In
addition to these static characteristics we studied non-equilibrium properties
of the system such as the boundary condition (i.e., hydrodynamic interface
position and slip length) and the steady-state velocity of the droplet when a
bulk force is exerted on it. 

An analytical relation predicting the velocity field in the direction of the
flow is derived and the scaling  of the velocity of the droplet's center of
mass discussed. We show that there are two different regime in the droplet
dynamics: For small droplets, friction at the substrate dominates dissipation
and the droplets velocity, $U$, increases like $\sqrt{N}$ with the number of
particles, $N$, at constant acceleration.  For large droplets, however, the
viscous dissipation of the flow in the interior of the droplet dictates the
behavior and one observes $U \sim N$. Surprisingly, we find that the larger is
the contact angle, and thus the smaller is the area of contact, the larger are
the surface effects.  

While our study is restricted to cylindrical droplets for computational
constraints the qualitative arguments carry over to cap-shaped droplets. The
cross-over between the surface- and the volume-dominated dissipation behavior
occurs at a droplet radius on the order $R_c \sim \delta / (1-\cos \theta_E)$.
Using a slip length on the order of micrometer as extracted from recent
dewetting experiments by Fetzer {\em at al.}\cite{Fetzer05,Fetzer06} on
low-molecular-weight polystyrene melts on octadecyl- (OTS) and
dodecyltrichlorosilane (DTS) polymer brushes and a contact angle of $\Theta_E
\approx 67^o$ we predict $R_c \approx 3 \mu$m which is in an experimentally
accessible range. 

\begin{acknowledgments}
The authors thank C.~Pastorino for fruitful discussions and the NIC
computing center at J{\"u}lich for the computational resources. This 
research is financially supported by the DFG priority program ``nano-
and microfluidics'' under grant Mu1674/3.
\end{acknowledgments} 

%\bibliography{Bibliography}
\bibliographystyle{apsrev}
\bibliography{bibtex}

\begin{thebibliography}{58}
\expandafter\ifx\csname natexlab\endcsname\relax\def\natexlab#1{#1}\fi
\expandafter\ifx\csname bibnamefont\endcsname\relax
  \def\bibnamefont#1{#1}\fi
\expandafter\ifx\csname bibfnamefont\endcsname\relax
  \def\bibfnamefont#1{#1}\fi
\expandafter\ifx\csname citenamefont\endcsname\relax
  \def\citenamefont#1{#1}\fi
\expandafter\ifx\csname url\endcsname\relax
  \def\url#1{\texttt{#1}}\fi
\expandafter\ifx\csname urlprefix\endcsname\relax\def\urlprefix{URL }\fi
\providecommand{\bibinfo}[2]{#2}
\providecommand{\eprint}[2][]{\url{#2}}

\bibitem[{\citenamefont{Squires and Quake}(2005)}]{Quake2005}
\bibinfo{author}{\bibfnamefont{T.~M.} \bibnamefont{Squires}} \bibnamefont{and}
  \bibinfo{author}{\bibfnamefont{S.~R.} \bibnamefont{Quake}},
  \bibinfo{journal}{Rev. Mod. Phys.} \textbf{\bibinfo{volume}{77}},
  \bibinfo{pages}{977} (\bibinfo{year}{2005}).

\bibitem[{\citenamefont{Velarde}(1998)}]{Velarde1998}
\bibinfo{author}{\bibfnamefont{M.~G.} \bibnamefont{Velarde}},
  \bibinfo{journal}{Philos. Trans. R. Soc. London A}
  \textbf{\bibinfo{volume}{356}}, \bibinfo{pages}{829} (\bibinfo{year}{1998}).

\bibitem[{\citenamefont{Brochard}(1989)}]{Brochard1989}
\bibinfo{author}{\bibfnamefont{F.}~\bibnamefont{Brochard}},
  \bibinfo{journal}{Langmuir} \textbf{\bibinfo{volume}{5}},
  \bibinfo{pages}{432} (\bibinfo{year}{1989}).

\bibitem[{\citenamefont{Lee et~al.}(2002)\citenamefont{Lee, Kwok, and
  Laibinis}}]{Laibinis2002}
\bibinfo{author}{\bibfnamefont{S.-W.} \bibnamefont{Lee}},
  \bibinfo{author}{\bibfnamefont{D.~Y.} \bibnamefont{Kwok}}, \bibnamefont{and}
  \bibinfo{author}{\bibfnamefont{P.~E.} \bibnamefont{Laibinis}},
  \bibinfo{journal}{Phys. Rev. E} \textbf{\bibinfo{volume}{65}},
  \bibinfo{pages}{051602} (\bibinfo{year}{2002}).

\bibitem[{\citenamefont{Santos and Ondar{\c c}uhu}(1995)}]{Ondarcuhu1995}
\bibinfo{author}{\bibfnamefont{F.~D.~D.} \bibnamefont{Santos}}
  \bibnamefont{and} \bibinfo{author}{\bibfnamefont{T.}~\bibnamefont{Ondar{\c
  c}uhu}}, \bibinfo{journal}{Phys. Rev. Lett.} \textbf{\bibinfo{volume}{75}},
  \bibinfo{pages}{2972} (\bibinfo{year}{1995}).

\bibitem[{\citenamefont{Thiele et~al.}(2004)\citenamefont{Thiele, John, and
  B{\"a}r}}]{Thiele2004}
\bibinfo{author}{\bibfnamefont{U.}~\bibnamefont{Thiele}},
  \bibinfo{author}{\bibfnamefont{K.}~\bibnamefont{John}}, \bibnamefont{and}
  \bibinfo{author}{\bibfnamefont{M.}~\bibnamefont{B{\"a}r}},
  \bibinfo{journal}{Phys. Rev. Lett.} \textbf{\bibinfo{volume}{93}},
  \bibinfo{pages}{027802} (\bibinfo{year}{2004}).

\bibitem[{\citenamefont{Rauscher et~al.}(2007)\citenamefont{Rauscher, Dietrich,
  and Koplik}}]{Rauscher07}
\bibinfo{author}{\bibfnamefont{M.}~\bibnamefont{Rauscher}},
  \bibinfo{author}{\bibfnamefont{S.}~\bibnamefont{Dietrich}}, \bibnamefont{and}
  \bibinfo{author}{\bibfnamefont{J.}~\bibnamefont{Koplik}},
  \bibinfo{journal}{Phys. Rev. Lett.} \textbf{\bibinfo{volume}{98}},
  \bibinfo{pages}{224504} (\bibinfo{year}{2007}).

\bibitem[{\citenamefont{Mahadevan and Pomeau}(1999)}]{Pomeau1999}
\bibinfo{author}{\bibfnamefont{L.}~\bibnamefont{Mahadevan}} \bibnamefont{and}
  \bibinfo{author}{\bibfnamefont{Y.}~\bibnamefont{Pomeau}},
  \bibinfo{journal}{Phys. Fluids} \textbf{\bibinfo{volume}{11}},
  \bibinfo{pages}{2449} (\bibinfo{year}{1999}).

\bibitem[{\citenamefont{Thiele et~al.}(2001)\citenamefont{Thiele, Velarde,
  Neuffer, Bestehorn, and Pomeau}}]{Pomeau2001}
\bibinfo{author}{\bibfnamefont{U.}~\bibnamefont{Thiele}},
  \bibinfo{author}{\bibfnamefont{M.~G.} \bibnamefont{Velarde}},
  \bibinfo{author}{\bibfnamefont{K.}~\bibnamefont{Neuffer}},
  \bibinfo{author}{\bibfnamefont{M.}~\bibnamefont{Bestehorn}},
  \bibnamefont{and} \bibinfo{author}{\bibfnamefont{Y.}~\bibnamefont{Pomeau}},
  \bibinfo{journal}{Phys. Rev. E} \textbf{\bibinfo{volume}{64}},
  \bibinfo{pages}{061601} (\bibinfo{year}{2001}).

\bibitem[{\citenamefont{Thiele et~al.}(2002)\citenamefont{Thiele, Neuffer,
  Bestehorn, Pomeau, and Velarde}}]{Thiele02}
\bibinfo{author}{\bibfnamefont{U.}~\bibnamefont{Thiele}},
  \bibinfo{author}{\bibfnamefont{K.}~\bibnamefont{Neuffer}},
  \bibinfo{author}{\bibfnamefont{M.}~\bibnamefont{Bestehorn}},
  \bibinfo{author}{\bibfnamefont{Y.}~\bibnamefont{Pomeau}}, \bibnamefont{and}
  \bibinfo{author}{\bibfnamefont{M.~G.} \bibnamefont{Velarde}},
  \bibinfo{journal}{Colloids And Surfaces A-Physicochemical And Engineering
  Aspects} \textbf{\bibinfo{volume}{206}}, \bibinfo{pages}{87}
  (\bibinfo{year}{2002}).

\bibitem[{\citenamefont{Huppert}(1982)}]{Huppert82}
\bibinfo{author}{\bibfnamefont{H.~E.} \bibnamefont{Huppert}},
  \bibinfo{journal}{Nature} \textbf{\bibinfo{volume}{300}},
  \bibinfo{pages}{427} (\bibinfo{year}{1982}).

\bibitem[{\citenamefont{Rio et~al.}(2005)\citenamefont{Rio, Daerr, Andreotti,
  and Limat}}]{Rio05}
\bibinfo{author}{\bibfnamefont{E.}~\bibnamefont{Rio}},
  \bibinfo{author}{\bibfnamefont{A.}~\bibnamefont{Daerr}},
  \bibinfo{author}{\bibfnamefont{B.}~\bibnamefont{Andreotti}},
  \bibnamefont{and} \bibinfo{author}{\bibfnamefont{L.}~\bibnamefont{Limat}},
  \bibinfo{journal}{Phys. Rev. Lett.} \textbf{\bibinfo{volume}{94}},
  \bibinfo{pages}{024503} (\bibinfo{year}{2005}).

\bibitem[{\citenamefont{Grand et~al.}(2005)\citenamefont{Grand, Daerr, and
  Limat}}]{LeGrand05}
\bibinfo{author}{\bibfnamefont{N.~L.} \bibnamefont{Grand}},
  \bibinfo{author}{\bibfnamefont{A.}~\bibnamefont{Daerr}}, \bibnamefont{and}
  \bibinfo{author}{\bibfnamefont{L.}~\bibnamefont{Limat}},
  \bibinfo{journal}{Journal Of Fluid Mechanics} \textbf{\bibinfo{volume}{541}},
  \bibinfo{pages}{293} (\bibinfo{year}{2005}).

\bibitem[{\citenamefont{Grand-Piteira et~al.}(2006)\citenamefont{Grand-Piteira,
  Daerr, and Limat}}]{LeGrand06b}
\bibinfo{author}{\bibfnamefont{N.~L.} \bibnamefont{Grand-Piteira}},
  \bibinfo{author}{\bibfnamefont{A.}~\bibnamefont{Daerr}}, \bibnamefont{and}
  \bibinfo{author}{\bibfnamefont{L.}~\bibnamefont{Limat}},
  \bibinfo{journal}{Phys. Rev. Lett.} \textbf{\bibinfo{volume}{96}},
  \bibinfo{pages}{254503} (\bibinfo{year}{2006}).

\bibitem[{\citenamefont{Kim et~al.}(2002{\natexlab{a}})\citenamefont{Kim, Lee,
  and Kang}}]{Kim02d}
\bibinfo{author}{\bibfnamefont{H.~Y.} \bibnamefont{Kim}},
  \bibinfo{author}{\bibfnamefont{H.~J.} \bibnamefont{Lee}}, \bibnamefont{and}
  \bibinfo{author}{\bibfnamefont{B.~H.} \bibnamefont{Kang}},
  \bibinfo{journal}{Journal Of Colloid And Interface Science}
  \textbf{\bibinfo{volume}{247}}, \bibinfo{pages}{372}
  (\bibinfo{year}{2002}{\natexlab{a}}).

\bibitem[{\citenamefont{Brochard-Wyart and de~Gennes}(1992)}]{BrochardF1992}
\bibinfo{author}{\bibfnamefont{F.}~\bibnamefont{Brochard-Wyart}}
  \bibnamefont{and} \bibinfo{author}{\bibfnamefont{P.~G.}
  \bibnamefont{de~Gennes}}, \bibinfo{journal}{Adv. Coll. Interf. Sci.}
  \textbf{\bibinfo{volume}{39}}, \bibinfo{pages}{1} (\bibinfo{year}{1992}).

\bibitem[{\citenamefont{Tanner}(1979)}]{Tanner79}
\bibinfo{author}{\bibfnamefont{L.~H.} \bibnamefont{Tanner}},
  \bibinfo{journal}{Journal Of Physics D-Applied Physics}
  \textbf{\bibinfo{volume}{12}}, \bibinfo{pages}{1473} (\bibinfo{year}{1979}).

\bibitem[{\citenamefont{de~Ruijter et~al.}(1997)\citenamefont{de~Ruijter,
  de~Coninck, Blake, Clarke, and Rankin}}]{Ruijter97}
\bibinfo{author}{\bibfnamefont{M.~J.} \bibnamefont{de~Ruijter}},
  \bibinfo{author}{\bibfnamefont{J.}~\bibnamefont{de~Coninck}},
  \bibinfo{author}{\bibfnamefont{T.~D.} \bibnamefont{Blake}},
  \bibinfo{author}{\bibfnamefont{A.}~\bibnamefont{Clarke}}, \bibnamefont{and}
  \bibinfo{author}{\bibfnamefont{A.}~\bibnamefont{Rankin}},
  \bibinfo{journal}{Langmuir} \textbf{\bibinfo{volume}{13}},
  \bibinfo{pages}{7293} (\bibinfo{year}{1997}).

\bibitem[{\citenamefont{de~Ruijter et~al.}(1999)\citenamefont{de~Ruijter,
  Blake, and de~Coninck}}]{Ruijter99}
\bibinfo{author}{\bibfnamefont{M.~J.} \bibnamefont{de~Ruijter}},
  \bibinfo{author}{\bibfnamefont{T.~D.} \bibnamefont{Blake}}, \bibnamefont{and}
  \bibinfo{author}{\bibfnamefont{J.}~\bibnamefont{de~Coninck}},
  \bibinfo{journal}{Langmuir} \textbf{\bibinfo{volume}{15}},
  \bibinfo{pages}{7836} (\bibinfo{year}{1999}).

\bibitem[{\citenamefont{Milchev and Binder}(2002)}]{Milchev02e}
\bibinfo{author}{\bibfnamefont{A.}~\bibnamefont{Milchev}} \bibnamefont{and}
  \bibinfo{author}{\bibfnamefont{K.}~\bibnamefont{Binder}},
  \bibinfo{journal}{J. Chem. Phys.} \textbf{\bibinfo{volume}{116}},
  \bibinfo{pages}{7691} (\bibinfo{year}{2002}).

\bibitem[{\citenamefont{Heine et~al.}(2003)\citenamefont{Heine, Grest, and
  Webb}}]{Heine03}
\bibinfo{author}{\bibfnamefont{D.~R.} \bibnamefont{Heine}},
  \bibinfo{author}{\bibfnamefont{G.~S.} \bibnamefont{Grest}}, \bibnamefont{and}
  \bibinfo{author}{\bibfnamefont{E.~B.} \bibnamefont{Webb}},
  \bibinfo{journal}{Phys. Rev. E} \textbf{\bibinfo{volume}{68}},
  \bibinfo{pages}{061603} (\bibinfo{year}{2003}).

\bibitem[{\citenamefont{Heine et~al.}(2004)\citenamefont{Heine, Grest, and
  Webb}}]{Heine04}
\bibinfo{author}{\bibfnamefont{D.~R.} \bibnamefont{Heine}},
  \bibinfo{author}{\bibfnamefont{G.~S.} \bibnamefont{Grest}}, \bibnamefont{and}
  \bibinfo{author}{\bibfnamefont{E.~B.} \bibnamefont{Webb}},
  \bibinfo{journal}{Phys. Rev. E} \textbf{\bibinfo{volume}{70}},
  \bibinfo{pages}{011606} (\bibinfo{year}{2004}).

\bibitem[{\citenamefont{Heine et~al.}(2005{\natexlab{a}})\citenamefont{Heine,
  Grest, and Webb}}]{Webb2005}
\bibinfo{author}{\bibfnamefont{D.~R.} \bibnamefont{Heine}},
  \bibinfo{author}{\bibfnamefont{G.~S.} \bibnamefont{Grest}}, \bibnamefont{and}
  \bibinfo{author}{\bibfnamefont{E.~B.} \bibnamefont{Webb}},
  \bibinfo{journal}{Phys. Rev. Lett.} \textbf{\bibinfo{volume}{95}},
  \bibinfo{pages}{107801} (\bibinfo{year}{2005}{\natexlab{a}}).

\bibitem[{\citenamefont{Heine et~al.}(2005{\natexlab{b}})\citenamefont{Heine,
  Grest, and Webb}}]{Heine05b}
\bibinfo{author}{\bibfnamefont{D.~R.} \bibnamefont{Heine}},
  \bibinfo{author}{\bibfnamefont{G.~S.} \bibnamefont{Grest}}, \bibnamefont{and}
  \bibinfo{author}{\bibfnamefont{E.~B.} \bibnamefont{Webb}},
  \bibinfo{journal}{Langmuir} \textbf{\bibinfo{volume}{21}},
  \bibinfo{pages}{7959} (\bibinfo{year}{2005}{\natexlab{b}}).

\bibitem[{\citenamefont{{G. S. Grest } and {K. Kremer}}(1990)}]{kremer_grest}
\bibinfo{author}{\bibnamefont{{G. S. Grest }}} \bibnamefont{and}
  \bibinfo{author}{\bibnamefont{{K. Kremer}}}, \bibinfo{journal}{Phys. Rev. A}
  \textbf{\bibinfo{volume}{33}}, \bibinfo{pages}{3628} (\bibinfo{year}{1990}).

\bibitem[{\citenamefont{Bennemann et~al.}(1999)\citenamefont{Bennemann, Paul,
  Baschnagel, and Binder}}]{Bennemann99d}
\bibinfo{author}{\bibfnamefont{C.}~\bibnamefont{Bennemann}},
  \bibinfo{author}{\bibfnamefont{W.}~\bibnamefont{Paul}},
  \bibinfo{author}{\bibfnamefont{J.}~\bibnamefont{Baschnagel}},
  \bibnamefont{and} \bibinfo{author}{\bibfnamefont{K.}~\bibnamefont{Binder}},
  \bibinfo{journal}{J. Phys.: Condens. Matter} \textbf{\bibinfo{volume}{11}},
  \bibinfo{pages}{2179} (\bibinfo{year}{1999}).

\bibitem[{\citenamefont{M{\"u}ller and MacDowell}(2000)}]{Muller00f}
\bibinfo{author}{\bibfnamefont{M.}~\bibnamefont{M{\"u}ller}} \bibnamefont{and}
  \bibinfo{author}{\bibfnamefont{L.~G.} \bibnamefont{MacDowell}},
  \bibinfo{journal}{Macromolecules} \textbf{\bibinfo{volume}{33}},
  \bibinfo{pages}{3902} (\bibinfo{year}{2000}).

\bibitem[{\citenamefont{M{\"u}ller et~al.}(2003)\citenamefont{M{\"u}ller,
  MacDowell, and Yethiraj}}]{Muller03d}
\bibinfo{author}{\bibfnamefont{M.}~\bibnamefont{M{\"u}ller}},
  \bibinfo{author}{\bibfnamefont{L.~G.} \bibnamefont{MacDowell}},
  \bibnamefont{and} \bibinfo{author}{\bibfnamefont{A.}~\bibnamefont{Yethiraj}},
  \bibinfo{journal}{J. Chem. Phys.} \textbf{\bibinfo{volume}{118}},
  \bibinfo{pages}{2929} (\bibinfo{year}{2003}).

\bibitem[{\citenamefont{Pastorino et~al.}(2006)\citenamefont{Pastorino, Binder,
  Kreer, and M{\"u}ller}}]{Pastorino2006}
\bibinfo{author}{\bibfnamefont{C.}~\bibnamefont{Pastorino}},
  \bibinfo{author}{\bibfnamefont{K.}~\bibnamefont{Binder}},
  \bibinfo{author}{\bibfnamefont{T.}~\bibnamefont{Kreer}}, \bibnamefont{and}
  \bibinfo{author}{\bibfnamefont{M.}~\bibnamefont{M{\"u}ller}},
  \bibinfo{journal}{J. Chem. Phys.} \textbf{\bibinfo{volume}{124}},
  \bibinfo{pages}{064902} (\bibinfo{year}{2006}).

\bibitem[{\citenamefont{Hoogerbrugge and Koelman}(1992)}]{Koelman1992}
\bibinfo{author}{\bibfnamefont{P.~J.} \bibnamefont{Hoogerbrugge}}
  \bibnamefont{and} \bibinfo{author}{\bibfnamefont{J.~M. V.~A.}
  \bibnamefont{Koelman}}, \bibinfo{journal}{Europhys. Lett.}
  \textbf{\bibinfo{volume}{19}}, \bibinfo{pages}{155} (\bibinfo{year}{1992}).

\bibitem[{\citenamefont{Espa{\~n}ol and Warren}(1995)}]{Warren1995}
\bibinfo{author}{\bibfnamefont{P.}~\bibnamefont{Espa{\~n}ol}} \bibnamefont{and}
  \bibinfo{author}{\bibfnamefont{P.}~\bibnamefont{Warren}},
  \bibinfo{journal}{Europhys. Lett.} \textbf{\bibinfo{volume}{30}},
  \bibinfo{pages}{191} (\bibinfo{year}{1995}).

\bibitem[{\citenamefont{Soddemann et~al.}(2003)\citenamefont{Soddemann,
  D{\"u}nweg, and Kremer}}]{Kremer2003}
\bibinfo{author}{\bibfnamefont{T.}~\bibnamefont{Soddemann}},
  \bibinfo{author}{\bibfnamefont{B.}~\bibnamefont{D{\"u}nweg}},
  \bibnamefont{and} \bibinfo{author}{\bibfnamefont{K.}~\bibnamefont{Kremer}},
  \bibinfo{journal}{Phys. Rev. E} \textbf{\bibinfo{volume}{68}},
  \bibinfo{pages}{046702} (\bibinfo{year}{2003}).

\bibitem[{\citenamefont{D{\"u}nweg and Paul}(1991)}]{Dunweg1991}
\bibinfo{author}{\bibfnamefont{B.}~\bibnamefont{D{\"u}nweg}} \bibnamefont{and}
  \bibinfo{author}{\bibfnamefont{W.}~\bibnamefont{Paul}},
  \bibinfo{journal}{Int. J. Mod. Phys. C} \textbf{\bibinfo{volume}{2}},
  \bibinfo{pages}{817} (\bibinfo{year}{1991}).

\bibitem[{\citenamefont{MacDowell et~al.}(2000)\citenamefont{MacDowell,
  M{\"u}ller, Vega, and Binder}}]{MacDowell00}
\bibinfo{author}{\bibfnamefont{L.~G.} \bibnamefont{MacDowell}},
  \bibinfo{author}{\bibfnamefont{M.}~\bibnamefont{M{\"u}ller}},
  \bibinfo{author}{\bibfnamefont{C.}~\bibnamefont{Vega}}, \bibnamefont{and}
  \bibinfo{author}{\bibfnamefont{K.}~\bibnamefont{Binder}},
  \bibinfo{journal}{J. Chem. Phys.} \textbf{\bibinfo{volume}{113}},
  \bibinfo{pages}{419} (\bibinfo{year}{2000}).

\bibitem[{\citenamefont{Doi and Edwards}(1996)}]{DoiEdwards}
\bibinfo{author}{\bibfnamefont{M.}~\bibnamefont{Doi}} \bibnamefont{and}
  \bibinfo{author}{\bibfnamefont{S.~F.} \bibnamefont{Edwards}},
  \emph{\bibinfo{title}{The Theory of Polymer Dynamics}}
  (\bibinfo{publisher}{Clarendon Press}, \bibinfo{address}{Oxford},
  \bibinfo{year}{1996}).

\bibitem[{\citenamefont{McQuarrie}(2000)}]{McQuarrie}
\bibinfo{author}{\bibfnamefont{D.}~\bibnamefont{McQuarrie}},
  \emph{\bibinfo{title}{Statistical Mechanics}} (\bibinfo{publisher}{University
  Science Books}, \bibinfo{address}{Sausalito}, \bibinfo{year}{2000}).

\bibitem[{\citenamefont{Nijmeijer et~al.}(1988)\citenamefont{Nijmeijer, Bakker,
  Bruin, and Sikkenk}}]{Sikkenk1988}
\bibinfo{author}{\bibfnamefont{M.~J.~P.} \bibnamefont{Nijmeijer}},
  \bibinfo{author}{\bibfnamefont{A.~F.} \bibnamefont{Bakker}},
  \bibinfo{author}{\bibfnamefont{C.}~\bibnamefont{Bruin}}, \bibnamefont{and}
  \bibinfo{author}{\bibfnamefont{J.~H.} \bibnamefont{Sikkenk}},
  \bibinfo{journal}{J. Chem. Phys.} \textbf{\bibinfo{volume}{89}},
  \bibinfo{pages}{3789} (\bibinfo{year}{1988}).

\bibitem[{\citenamefont{Orea et~al.}(2002)\citenamefont{Orea, Duda, and
  Alejandre}}]{Alejandre2002}
\bibinfo{author}{\bibfnamefont{P.}~\bibnamefont{Orea}},
  \bibinfo{author}{\bibfnamefont{Y.}~\bibnamefont{Duda}}, \bibnamefont{and}
  \bibinfo{author}{\bibfnamefont{J.}~\bibnamefont{Alejandre}},
  \bibinfo{journal}{J. Chem. Phys.} \textbf{\bibinfo{volume}{118}},
  \bibinfo{pages}{5635} (\bibinfo{year}{2002}).

\bibitem[{\citenamefont{Hienola et~al.}(2007)\citenamefont{Hienola, Winkler,
  Wagner, Vehkamäki, Lauri, Napari, and Kulmala}}]{Kulmala2007}
\bibinfo{author}{\bibfnamefont{A.~I.} \bibnamefont{Hienola}},
  \bibinfo{author}{\bibfnamefont{P.~M.} \bibnamefont{Winkler}},
  \bibinfo{author}{\bibfnamefont{P.~E.} \bibnamefont{Wagner}},
  \bibinfo{author}{\bibfnamefont{H.}~\bibnamefont{Vehkamäki}},
  \bibinfo{author}{\bibfnamefont{A.}~\bibnamefont{Lauri}},
  \bibinfo{author}{\bibfnamefont{I.}~\bibnamefont{Napari}}, \bibnamefont{and}
  \bibinfo{author}{\bibfnamefont{M.}~\bibnamefont{Kulmala}},
  \bibinfo{journal}{J. Chem. Phys.} \textbf{\bibinfo{volume}{126}},
  \bibinfo{pages}{094705} (\bibinfo{year}{2007}).

\bibitem[{\citenamefont{Amirfazlia et~al.}(1998)\citenamefont{Amirfazlia,
  Kwoka, Gaydosb, and Neumanna}}]{Neumanna1998}
\bibinfo{author}{\bibfnamefont{A.}~\bibnamefont{Amirfazlia}},
  \bibinfo{author}{\bibfnamefont{D.~Y.} \bibnamefont{Kwoka}},
  \bibinfo{author}{\bibfnamefont{J.}~\bibnamefont{Gaydosb}}, \bibnamefont{and}
  \bibinfo{author}{\bibfnamefont{A.~W.} \bibnamefont{Neumanna}},
  \bibinfo{journal}{J. Colloid Int. Sci.} \textbf{\bibinfo{volume}{205}},
  \bibinfo{pages}{1} (\bibinfo{year}{1998}).

\bibitem[{\citenamefont{Young}(1805)}]{Young}
\bibinfo{author}{\bibfnamefont{T.}~\bibnamefont{Young}},
  \bibinfo{journal}{Philos.\ Trans.\ R.\ Soc.\ London}
  \textbf{\bibinfo{volume}{5}}, \bibinfo{pages}{65} (\bibinfo{year}{1805}).

\bibitem[{\citenamefont{de~Gennes}(1985)}]{deGennes1985}
\bibinfo{author}{\bibfnamefont{P.~G.} \bibnamefont{de~Gennes}},
  \bibinfo{journal}{Rev. Mod. Phys.} \textbf{\bibinfo{volume}{57}},
  \bibinfo{pages}{827} (\bibinfo{year}{1985}).

\bibitem[{\citenamefont{Adams and Henderson}(1991)}]{Adams91}
\bibinfo{author}{\bibfnamefont{P.}~\bibnamefont{Adams}} \bibnamefont{and}
  \bibinfo{author}{\bibfnamefont{J.~R.} \bibnamefont{Henderson}},
  \bibinfo{journal}{Molecular Physics} \textbf{\bibinfo{volume}{73}},
  \bibinfo{pages}{1383} (\bibinfo{year}{1991}).

\bibitem[{\citenamefont{Vanswol and Henderson}(1991)}]{Vanswol91}
\bibinfo{author}{\bibfnamefont{F.}~\bibnamefont{Vanswol}} \bibnamefont{and}
  \bibinfo{author}{\bibfnamefont{J.~R.} \bibnamefont{Henderson}},
  \bibinfo{journal}{Phys. Rev. A} \textbf{\bibinfo{volume}{43}},
  \bibinfo{pages}{2932} (\bibinfo{year}{1991}).

\bibitem[{\citenamefont{Bruin et~al.}(1995)\citenamefont{Bruin, Nijmeijer, and
  Crevecoeur}}]{Bruin95}
\bibinfo{author}{\bibfnamefont{C.}~\bibnamefont{Bruin}},
  \bibinfo{author}{\bibfnamefont{M.~J.~P.} \bibnamefont{Nijmeijer}},
  \bibnamefont{and} \bibinfo{author}{\bibfnamefont{R.~M.}
  \bibnamefont{Crevecoeur}}, \bibinfo{journal}{J. Chem. Phys.}
  \textbf{\bibinfo{volume}{102}}, \bibinfo{pages}{7622} (\bibinfo{year}{1995}).

\bibitem[{\citenamefont{Muller and Macdowell}(2000)}]{Muller00i}
\bibinfo{author}{\bibfnamefont{M.}~\bibnamefont{Muller}} \bibnamefont{and}
  \bibinfo{author}{\bibfnamefont{L.~G.} \bibnamefont{Macdowell}},
  \bibinfo{journal}{Macromolecules} \textbf{\bibinfo{volume}{33}},
  \bibinfo{pages}{3902} (\bibinfo{year}{2000}).

\bibitem[{\citenamefont{Bocquet and Barrat}(2007)}]{Bocquet07c}
\bibinfo{author}{\bibfnamefont{L.}~\bibnamefont{Bocquet}} \bibnamefont{and}
  \bibinfo{author}{\bibfnamefont{J.~L.} \bibnamefont{Barrat}},
  \bibinfo{journal}{Soft Matter} \textbf{\bibinfo{volume}{3}},
  \bibinfo{pages}{685} (\bibinfo{year}{2007}).

\bibitem[{\citenamefont{Lauga et~al.}(2007)\citenamefont{Lauga, Brenner, and
  Stone}}]{Lauga07}
\bibinfo{author}{\bibfnamefont{E.}~\bibnamefont{Lauga}},
  \bibinfo{author}{\bibfnamefont{M.~P.} \bibnamefont{Brenner}},
  \bibnamefont{and} \bibinfo{author}{\bibfnamefont{H.~A.} \bibnamefont{Stone}},
  \emph{\bibinfo{title}{The no-slip boundary condition, Chapt 15 in
  Experimental Fluid Dynamics edited by J. Foss, C. Tropea and A. Yarin,
  cond-mat/0501557}} (\bibinfo{publisher}{Springer}, \bibinfo{address}{Berlin},
  \bibinfo{year}{2007}).

\bibitem[{\citenamefont{Thompson and Troian}(1997)}]{Troian1997}
\bibinfo{author}{\bibfnamefont{P.~A.} \bibnamefont{Thompson}} \bibnamefont{and}
  \bibinfo{author}{\bibfnamefont{S.~M.} \bibnamefont{Troian}},
  \bibinfo{journal}{Nature} \textbf{\bibinfo{volume}{389}},
  \bibinfo{pages}{360} (\bibinfo{year}{1997}).

\bibitem[{\citenamefont{Barrat and Bocquet}(1999)}]{Bocquet1999}
\bibinfo{author}{\bibfnamefont{J.-L.} \bibnamefont{Barrat}} \bibnamefont{and}
  \bibinfo{author}{\bibfnamefont{L.}~\bibnamefont{Bocquet}},
  \bibinfo{journal}{Phys. Rev. Lett.} \textbf{\bibinfo{volume}{82}},
  \bibinfo{pages}{4671} (\bibinfo{year}{1999}).

\bibitem[{\citenamefont{Navier}(1823)}]{Navier1823}
\bibinfo{author}{\bibfnamefont{C.~L. M.~H.} \bibnamefont{Navier}},
  \bibinfo{journal}{Mem. Roy. Sci. Inst. France} \textbf{\bibinfo{volume}{6}},
  \bibinfo{pages}{389} (\bibinfo{year}{1823}).

\bibitem[{\citenamefont{Bocquet and Barrat}(1993)}]{Bocquet1993}
\bibinfo{author}{\bibfnamefont{L.}~\bibnamefont{Bocquet}} \bibnamefont{and}
  \bibinfo{author}{\bibfnamefont{J.-L.} \bibnamefont{Barrat}},
  \bibinfo{journal}{Phys. Rev. Lett.} \textbf{\bibinfo{volume}{70}},
  \bibinfo{pages}{2726} (\bibinfo{year}{1993}).

\bibitem[{\citenamefont{Bocquet and Barrat}(1994)}]{Bocquet1994}
\bibinfo{author}{\bibfnamefont{L.}~\bibnamefont{Bocquet}} \bibnamefont{and}
  \bibinfo{author}{\bibfnamefont{J.-L.} \bibnamefont{Barrat}},
  \bibinfo{journal}{Phys. Rev. E} \textbf{\bibinfo{volume}{49}},
  \bibinfo{pages}{3079} (\bibinfo{year}{1994}).

\bibitem[{\citenamefont{Joly et~al.}(2006)\citenamefont{Joly, Ybert, and
  Bocquet}}]{Bocquet2006}
\bibinfo{author}{\bibfnamefont{L.}~\bibnamefont{Joly}},
  \bibinfo{author}{\bibfnamefont{C.}~\bibnamefont{Ybert}}, \bibnamefont{and}
  \bibinfo{author}{\bibfnamefont{L.}~\bibnamefont{Bocquet}},
  \bibinfo{journal}{Phys. Rev. Lett.} \textbf{\bibinfo{volume}{96}},
  \bibinfo{pages}{046101} (\bibinfo{year}{2006}).

\bibitem[{\citenamefont{Huh and Scriven}(1971)}]{Huh71}
\bibinfo{author}{\bibfnamefont{C.}~\bibnamefont{Huh}} \bibnamefont{and}
  \bibinfo{author}{\bibfnamefont{L.~E.} \bibnamefont{Scriven}},
  \bibinfo{journal}{Journal Of Colloid And Interface Science}
  \textbf{\bibinfo{volume}{35}}, \bibinfo{pages}{85} (\bibinfo{year}{1971}).

\bibitem[{\citenamefont{Kim et~al.}(2002{\natexlab{b}})\citenamefont{Kim,
  Faller, Yan, Abbott, and de~Pablo}}]{Kim02b}
\bibinfo{author}{\bibfnamefont{E.~B.} \bibnamefont{Kim}},
  \bibinfo{author}{\bibfnamefont{R.}~\bibnamefont{Faller}},
  \bibinfo{author}{\bibfnamefont{Q.}~\bibnamefont{Yan}},
  \bibinfo{author}{\bibfnamefont{N.~L.} \bibnamefont{Abbott}},
  \bibnamefont{and} \bibinfo{author}{\bibfnamefont{J.~J.}
  \bibnamefont{de~Pablo}}, \bibinfo{journal}{J. Chem. Phys.}
  \textbf{\bibinfo{volume}{117}}, \bibinfo{pages}{7781}
  (\bibinfo{year}{2002}{\natexlab{b}}).

\bibitem[{\citenamefont{Fetzer et~al.}(2005)\citenamefont{Fetzer, Jacobs,
  Munch, Wagner, and Witelski}}]{Fetzer05}
\bibinfo{author}{\bibfnamefont{R.}~\bibnamefont{Fetzer}},
  \bibinfo{author}{\bibfnamefont{K.}~\bibnamefont{Jacobs}},
  \bibinfo{author}{\bibfnamefont{A.}~\bibnamefont{Munch}},
  \bibinfo{author}{\bibfnamefont{B.}~\bibnamefont{Wagner}}, \bibnamefont{and}
  \bibinfo{author}{\bibfnamefont{T.~P.} \bibnamefont{Witelski}},
  \bibinfo{journal}{Phys. Rev. Lett.} \textbf{\bibinfo{volume}{95}},
  \bibinfo{pages}{127801} (\bibinfo{year}{2005}).

\bibitem[{\citenamefont{Fetzer et~al.}(2006)\citenamefont{Fetzer, Rauscher,
  Munch, Wagner, and Jacobs}}]{Fetzer06}
\bibinfo{author}{\bibfnamefont{R.}~\bibnamefont{Fetzer}},
  \bibinfo{author}{\bibfnamefont{M.}~\bibnamefont{Rauscher}},
  \bibinfo{author}{\bibfnamefont{A.}~\bibnamefont{Munch}},
  \bibinfo{author}{\bibfnamefont{B.~A.} \bibnamefont{Wagner}},
  \bibnamefont{and} \bibinfo{author}{\bibfnamefont{K.}~\bibnamefont{Jacobs}},
  \bibinfo{journal}{Europhys. Lett} \textbf{\bibinfo{volume}{75}},
  \bibinfo{pages}{638} (\bibinfo{year}{2006}).

\end{thebibliography}

\end{document}